\documentclass[journal]{IEEEtran}
\usepackage[colorlinks=true,bookmarks=false,citecolor=blue,urlcolor=blue]{hyperref} 

\makeatother
\usepackage{subfig}
\usepackage[table]{xcolor}
\usepackage{float}
\usepackage{caption}
\usepackage{float}
\usepackage{setspace}
\usepackage{graphicx}
\usepackage{graphicx,dblfloatfix}
\usepackage{tikz}
\usepackage{color}
\usepackage{natbib}
\usepackage{xcolor}
\usepackage{lipsum}
\usepackage{amsmath}
\usepackage{amssymb} 
\usetikzlibrary{shapes,arrows,shadows}
\ifCLASSINFOpdf
\else
\fi
\begin{document}
\title{Deep Neural Network Assisted Second-Order Perturbation-Based Nonlinearity Compensation}
%
%
%

\author{O. S. Sunish Kumar,
        Lutz Lampe, Shenghang Luo, Mrinmoy Jana, Jeebak Mitra, and Chuandong Li
        
\thanks{O. S. Sunish Kumar was with the Department of Electrical and Computer Engineering, The University of British Columbia, Vancouver, BC V6T 1Z4, Canada. He is now with the School of Engineering, Ulster University, Jordanstown, Northern Ireland, BT37 0QB, United Kingdom (e-mail: S.Orappanpara\_Soman@ulster.ac.uk).}
\thanks{Lutz Lampe, Shenghang Luo, and Mrinmoy Jana are with the Department of Electrical and Computer Engineering, The University of British Columbia, Vancouver, BC V6T 1Z4, Canada.}
\thanks{Jeebak Mitra and Chuandong Li are with the Huawei Technologies Canada, Ottawa, ON K2K 3J1, Canada.}}

%



\maketitle

\begin{abstract}
We propose a fiber nonlinearity post-compensation technique using the DNN and the second-order perturbation theory. We achieve $\sim1$ dB  $Q$-factor improvement for a 32 Gbaud PDM-64-QAM at 1200 km compared to the linear dispersion compensation.
\end{abstract}

\section{Introduction}
Fiber nonlinearity limits the transmission performance of the high baud-rate long-haul optical communication systems [1], [2]. Recently, model-based machine learning approaches have been considered for the fiber nonlinearity compensation (NLC) [3], for example, the neural network (NN)-assisted NLC (NN-NLC) [4]. The NN-NLC technique uses the first-order (FO) perturbation triplets as the input features to the deep neural network (DNN). That helps the DNN to effectively learn the underlying physical processes that govern the nonlinear interactions in the optical fiber [4]. Nowadays, there is an increasing thrust towards 800G systems using higher-order modulation formats. It should be noted that, in such systems, the FO perturbation approximation becomes inaccurate to model the nonlinear distortion field due to the increase in the transmit launch power. In other words, the higher-order perturbation terms become significant in such high data-rate systems. In this paper, unlike the NN-NLC technique, we propose using both FO triplets and second-order (SO) quintuples as the input features to the DNN to improve the NLC performance. This technique is referred to as the DNN-assisted SO-PB-NLC (DNN-SO-PB-NLC). 
\section{The DNN-SO-PB-NLC Technique}
The underlying premise of our proposed approach is that we add the SO perturbation term to the FO term in the series expansion-based approximation of the nonlinear distortion field to improve the approximation accuracy, thereby improving the NLC performance [5]. In conformance with theory, we propose to use an augmented DNN structure: one DNN to estimate the FO nonlinear distortion field and the other to estimate the SO distortion field. The estimated SO field is then added to the FO field. That improves the estimation accuracy of the overall nonlinear distortion field. 
\begin{figure}[h]
\centering\includegraphics[width=1\columnwidth,height=0.22\paperheight]{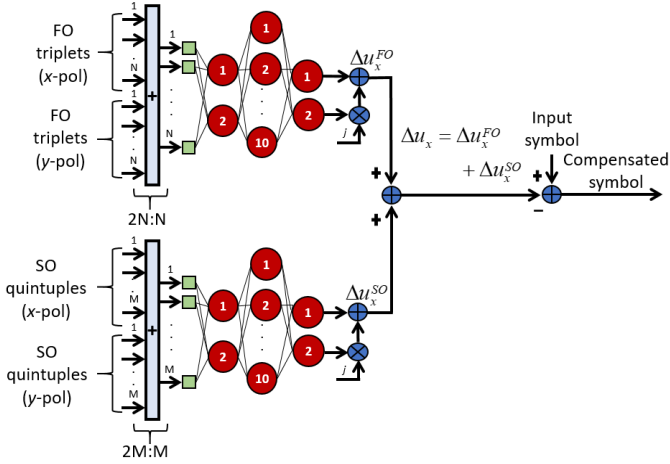}\caption{The DNN-SO-PB-NLC technique. $\Delta u_{x}^{\text{FO}}$ is the FO nonlinear distortion field.}
\end{figure}


Fig. 1 shows the network model of the DNN-SO-PB-NLC technique. The DNNs in Fig. 1 have 2 hidden layers consisting of 2 and 10 nodes, respectively. The output nodes provide the real and imaginary parts of the estimated FO/SO nonlinear distortion fields. In this technique, after generating $2N$ FO triplets, the intra- and inter-polarization triplets with the same indices are added to generate $N$ twin-triplets and given to the input layer of the upper DNN of Fig. 1. Similarly, $M$ twin-quintuples are generated from $2M$ SO quintuples and given to the input layer of the lower DNN in Fig. 1. In implementing the DNN-SO-PB-NLC technique, the estimated FO/SO nonlinear distortion field is subtracted from the symbol of interest to compensate for the nonlinearity effect. 

\section{Numerical Simulation and the Results}

Fig. 2 shows the numerical simulation setup for the DNN-SO-PB-NLC technique. The transmitter processing consists of the generation of single-channel 32 GBaud polarization-division multiplexed (PDM)-64-quadrature amplitude modulation (QAM) signal, root-raised cosine (RRC) pulse shaping filtering, 50$\%$ chromatic dispersion compensation (CDC), and the signal up-conversion using in-phase/quadrature-phase (I/Q) modulator. The up-converted signal is then transmitted through a 1200 km standard single-mode fiber (SSMF). The SSMF parameters considered are as given in [2]. An erbium-doped fiber amplifier (EDFA) with a gain of $16$ dB and a noise figure of 6 dB is used for the periodic amplification. At the receiver, after coherent detection, 50$\%$ CDC, RRC filter, and phase rotation, the proposed DNN-SO-PB-NLC technique is applied as post-compensation. 

\begin{figure}[h]
\centering\includegraphics[width=0.8\columnwidth,height=0.24\paperheight]{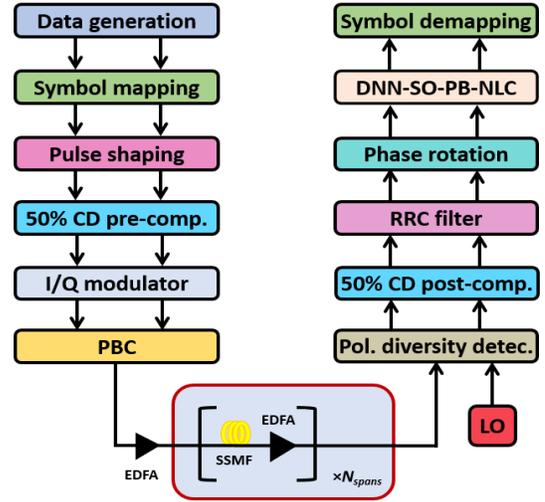}\caption{The simulation setup. PBC: polarization beam combiner, $N_{spans}$: number of spans, and LO: local oscillator.}
\end{figure}

\begin{figure}[h]
\includegraphics[width=1\columnwidth,height=0.22\paperheight]{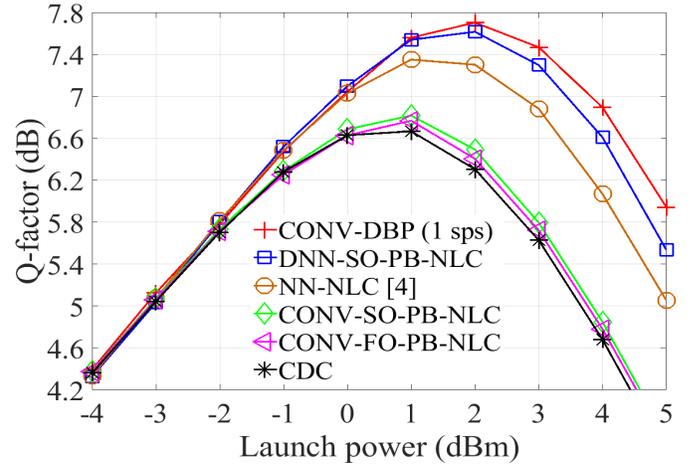}
\caption{Q-factor as a function of launch power. }
\end{figure}
\begin{figure}[h]
\includegraphics[width=1\columnwidth,height=0.22\paperheight]{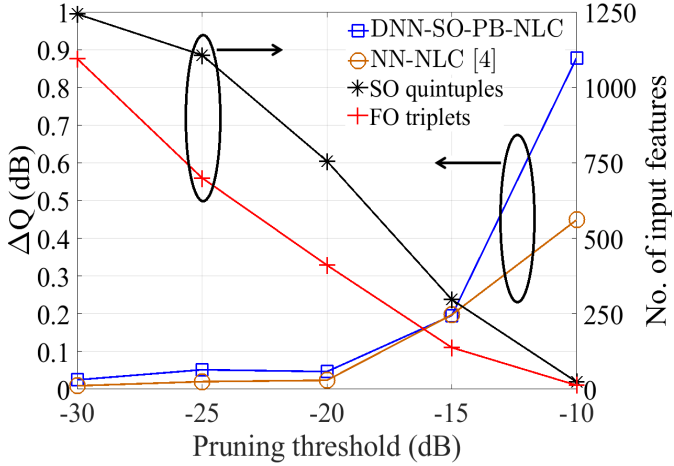}
\caption{$\Delta Q$ and the no. of input features as a function of the pruning threshold. sps: step/span.}
\end{figure}

It is evident from Fig. 3 that the DNN-SO-PB-NLC technique improves the $Q$-factor performance by $\sim0.27$ dB and $\sim1$ dB when compared to the NN-NLC and the CDC techniques, respectively. It also shows a significant performance improvement for the DNN-SO-PB-NLC technique when compared to the conventional (CONV) FO-/SO-PB-NLC techniques. To reduce the computational complexity, we trimmed off all the tensor weights that are less than a specified threshold value, and the network is re-trained and estimates the nonlinear distortion field. 

The results in Fig. 4 indicate that the pruning at the weight magnitude threshold of -15 dB results in 138 FO triplets and 268 SO quintuples, and the corresponding penalty $\Delta Q$, defined as the $Q$-factor difference between the case with no pruning and with pruning at the optimum launch power, is less than $0.2$ dB when compared to the case without pruning. Table 1 shows the computational complexity, and the results indicate that the real-valued multiplications per symbol for the proposed DNN-SO-PB-NLC is reduced by 23.75\% compared to the CONV digital back-propagation (DBP) technique. 

\begin{table}[tbph]
\begin{centering}
\caption{Computational complexity in terms of the number of real-valued multiplications/symbol.}
\vspace{-0.2cm}
\par\end{centering}
\centering{}%
\begin{tabular}{|p{8mm}|p{8mm}|p{8mm}|p{8mm}|p{8mm}|p{8mm}|p{8mm}|}
\hline 
\multicolumn{2}{|{c}|}{\centering{}\textbf{\scriptsize{}NN-NLC {[}4{]}}} & \multicolumn{2}{c|}{\textbf{\scriptsize{}DNN-SO-PB-NLC}} & \centering{\cellcolor{blue!25}} & \cellcolor{blue!25} & \cellcolor{blue!25}\tabularnewline
\hline 
\hline 
\multicolumn{2}{|c|}{\textbf{\scriptsize{}FO DNN part}} & \multicolumn{2}{c|}{\textbf{\scriptsize{}FO+SO DNN parts}} & \centering{\cellcolor{blue!25}} & \cellcolor{blue!25} & \cellcolor{blue!25}\tabularnewline
\hline 
\centering{}\textbf{\scriptsize{}DNN w/o prun.} & \centering{}\textbf{\scriptsize{}DNN w/ ~prun. @-15 dB} & \centering{}\textbf{\scriptsize{}DNN w/o prun.} & \centering{}\textbf{\scriptsize{}DNN w/ ~prun. @-15 dB} & \centering{}\textbf{\scriptsize{}CONV-FO-PB-NLC} & \centering{}\textbf{\scriptsize{}CONV-SO-PB-NLC} & \centering{}\textbf{\scriptsize{}CONV-DBP}\tabularnewline
\hline 
\centering{}{\scriptsize{}5,296} & \centering{}{\scriptsize{}592} & \centering{}{\scriptsize{}10,592} & \centering{}{\scriptsize{}1,704} & \centering{}{\scriptsize{}119} & \centering{}{\scriptsize{}1,547} & \centering{}{\scriptsize{}2,235}\tabularnewline
\hline 
\end{tabular}
\end{table}

\section{Conclusion}
We proposed a DNN-assisted SO-PB-NLC post-compensation technique to compensate for the fiber nonlinearity effect. We demonstrated that the proposed technique improves the $Q$-factor performance by $\sim0.27$ dB and $\sim1$ dB when compared to the NN-NLC and the CDC techniques, respectively. Further, we showed that the weight magnitude pruning of the DNN-SO-PB-NLC technique at a threshold of -15 dB reduces the number of real-valued multiplications per symbol by 531 when compared to the CONV-DBP technique.   

\end{document}